\documentclass[twocolumn,showpacs,prl,amsmath,amssymb]{revtex4}% Physical Review B

\usepackage{graphicx}% Include figure files
\usepackage{dcolumn}% Align table columns on decimal point
\usepackage{bm}% bold math

\newcommand{\be}{\begin{equation}}
\newcommand{\ee}{\end{equation}} 
\newcommand{\lb}{\label}

\newcommand{\bj}{{\mbox{\boldmath ${\j}$}}}
\newcommand{\btau}{{\mbox{\boldmath $\tau$}}}

\newcommand{\const}{({\rm const.})}

\newcommand{\ba}{{\bf a}}
\newcommand{\bb}{{\bf b}}

\newcommand{\br}{{\bf r}}
\newcommand{\bu}{{\bf u}}

\newcommand{\bx}{{\bf x}}
\newcommand{\by}{{\bf y}}
\newcommand{\bz}{{\bf z}}

\newcommand{\bB}{{\bf B}}

\newcommand{\bS}{{\bf S}}

\newcommand{\OL}{\overline}

\newcommand{\wt}{\widetilde}

\newcommand{\bepsilon}{{\mbox{\boldmath $\varepsilon$}}}

\newcommand{\grad}{{\mbox{\boldmath $\nabla$}}}
\newcommand{\bdot}{{\mbox{\boldmath $\cdot$}}}
\newcommand{\bdots}{{\mbox{\boldmath $:$}}}
\newcommand{\btimes}{{\mbox{\boldmath $\times$}}}

\newcommand{\boell}{{\mbox{\boldmath $\ell$}}}

\begin{document}

\preprint{APS/123-QED}

\title{Scale-Locality of Magnetohydrodynamic Turbulence}% Force line breaks with \\

\author{Hussein Aluie$^{1,2}$}
%\email{email1@jhu.edu}
\author{Gregory L. Eyink$^1$}%
%\email{email2.jhu.edu}
\affiliation{%
$^1$The Johns Hopkins University, Applied Mathematics \& Statistics, Baltimore, MD 21218, USA\\
$^2$Theoretical Division (T-5/CNLS), Los Alamos National Lab, Los Alamos, NM 87545, USA
}%

\date{\today}% It is always \today, today,

\begin{abstract}
We investigate the scale-locality of cascades of conserved invariants at high kinetic 
and magnetic Reynolds numbers in the ``inertial-inductive range'' of magnetohydrodynamic 
(MHD) turbulence, where velocity and magnetic field increments exhibit suitable power-law scaling. 
We prove that fluxes of total energy and cross-helicity---or, equivalently, fluxes of Els\"{a}sser  energies---are dominated by the contributions of local triads. Corresponding spectral transfers 
are also scale-local when defined using octave wavenumber bands. Flux and transfer of 
magnetic helicity may be dominated by non-local triads. The magnetic stretching term 
also may be dominated by non-local triads
but we prove that it can convert energy only  
between velocity and magnetic modes at comparable scales. We explain the disagreement
with numerical studies that have claimed conversion nonlocally between disparate scales.  
We present supporting data from a $1024^3$ simulation of forced MHD turbulence.

\end{abstract}

\pacs{95.30.Qd, 52.35.Ra, 47.27.Jv}
\maketitle

Magnetohydrodynamic (MHD) turbulence is pervasive in astrophysical systems.
Turbulent plasma fluctuations commonly possess power-law 
spectra over vast ranges of scales where both viscosity and resistivity are 
negligible. We call such ranges ``inertial-inductive'', since nonlinear dynamics 
(inertia/Lorentz force and convection/induction) dominates the physics at these scales. 
MHD plasma turbulence, with power-law scaling of both spectra and structure-functions in the 
inertial-inductive range, plays a central role in star formation, accretion of matter
near active galactic nuclei, solar physics, and in the generation of large-scale 
magnetic fields in such systems. There are several 
competing theories for the spectrum of strong MHD turbulence,  including those of 
Iroshnikov-Kraichnan \cite{Iroshnikov64, Kraichnan65}, Goldreich-Sridhar 
\cite{GoldreichSridhar95}, and Boldyrev
\cite{Boldyrev05}.  
All of these theories assume scale-locality of the nonlinear cascade, following the 
classical ideas of Richardson, Kolmogorov and Onsager for turbulence in neutral fluids.  
Scale-locality is fundamental to justify the universality of the postulated turbulent 
scaling laws. 

A consensus has been forming in recent years, however, that cascades in 
MHD turbulence are nonlocal processes  
\cite{Schekochihinetal04,Alexakisetal05a,Caratietal06,Yousefetal07}. 
Schekochihin et al. \cite{Schekochihinetal04} emphasized the
non-local nature of the interactions between the velocity and magnetic
fields as a hallmark of isotropic MHD turbulence. This conclusion was 
reaffirmed in several subsequent studies,  
most categorically by Yousef et al. 
\cite{Yousefetal07} who claimed that there is a direct exchange of energy 
between motions at the largest scales in the system, at which the flow is being 
stirred, and the magnetic field at arbitrarily small scales in the inductive range.
A more refined analysis of the locality of interactions was carried out by 
Alexakis et al. \cite{Alexakisetal05a} who concluded, based on direct 
numerical simulations (DNS) of MHD turbulence that the magnetic field gains 
energy at scales $\ell$ in the inertial range from the straining motions at all larger 
scales $>\ell$ and especially from the forcing scale $L\gg\ell$. Carati et al. 
\cite{Caratietal06} subsequently carried out DNS at a higher resolution,  
arriving at conclusions similar to \cite{Alexakisetal05a} and, furthermore, claiming
that there is non-local transfer of Els\"asser energies as well. Related ideas have 
surfaced in the accretion disk community \cite{FromangPapaloizou07,Fromangetal07}.

In this letter we address the scale-locality of MHD cascades by a direct analytical study 
of the equations:
$$  \partial_t\bu +(\bu\cdot\grad)\bu = -\grad p + 
        (\bb\cdot\grad)\bb + \nu \nabla^2 \bu + {\bf f} 
$$
\be  \partial_t\bb +(\bu\cdot\grad)\bb = (\bb\bdot\grad)\bu + \eta \nabla^2 \bb 
\lb{MHD-eq}\ee
for $\grad\cdot\bu=\grad\cdot\bb=0.$
Here $\bb=\bB/\sqrt{4\pi\rho}$ is the magnetic field in Alfv\'en velocity units and 
$p$ is total pressure (including magnetic pressure).  Our main conclusion is that,
under  very weak scaling assumptions,  MHD turbulence has scale-locality properties 
only a little less robust than those of hydrodynamic turbulence. We will support our 
analysis  with a pseudospectral DNS  at $1024^3$ resolution with phase-shift dealiasing. 
For our numerical work we choose viscosity $\nu$ and 
resistivity $\eta$ to be both equal to $1.1\times10^{-4}$. The external stirring force 
is a Taylor-Green flow ${\bf f} \equiv f_0 [ \sin(k_fx)\cos(k_f y)\cos(k_f z) 
{\hat \bx} - \cos(k_fx)\sin(k_fy)\cos(k_fz) {\hat \by}]$ applied at modes $k_f = 2$ 
with an amplitude $f_0 = 0.25$.  The Reynold's number based on the Taylor scale 
$\lambda_u = 2\pi  \sqrt{E_{u}}/ \big(\int dk k^2 E_u(k)\big)^{1/2}$ is $Re_{\lambda_u} 
= u_{rms}\lambda_u/\nu =909$.

Our proof of local cascade of invariants in MHD turbulence is very similar to that given
for hydrodynamic turbulence in \cite{Eyink05,EyinkAluie09,AluieEyink09}. We employ 
the spatial coarse-graining approach, commonly used as a modelling tool in the 
Large-Eddy Simulation (LES) community \cite{Germano92,MeneveauKatz00}. 
Coarse-grained fields are defined by $\OL f_\ell(\bx) = \int d\br\, G_\ell(\br) f(\bx+\br)$, 
with a filtering kernel $G_\ell(\br) = \ell^{-3}G(\br/\ell)$ which is sufficiently smooth and decays 
sufficiently rapidly for large $r$ \cite{Eyink05}. Coarse-grained MHD equations can then be 
written to describe $\OL\bu_\ell$ and $\OL\bb_\ell$, along with corresponding budgets 
for the quadratic invariants---energy, cross-helicity, and magnetic-helicity---at scales 
$\ge\ell$. See \cite{AluieEyink_long}. E.g. the time-derivative of large-scale energy 
$(1/2)[|\OL{\bu}_\ell|^2+|\OL{\bb}_\ell|^2],$ in addition to space-transport terms, contains
also as sink terms the {\it kinetic energy flux} $-\Pi_\ell^u=\grad\OL{\bu}_\ell:\btau_\ell$ and 
the {\it magnetic energy flux} $-\Pi_\ell^b=\OL{\bj}_\ell\bdot \bepsilon_\ell,$  with $\OL{\bj}_\ell
=\grad\btimes\OL{\bb}_\ell.$ Here $\tau_{\ell,\,ij}=\tau_\ell(u_i,u_j)-\tau_\ell(b_i,b_j)$ is 
the total stress generated by scales $<\ell,$ both the Reynolds stress and the Maxwell 
stress, and $\varepsilon_{\ell,\,i}=\epsilon_{ijk}\tau_\ell(u_j,b_k)$ is the electromotive 
force generated by scales $<\ell$.
We employ the notation
\be  \tau_\ell(f,g) =\OL{(fg)_\ell}-\OL{f}_\ell\OL{g}_\ell \lb{tau-def} \ee 
for the ``central moments'' of any fields $f(\bx),g(\bx)$ \cite{Germano92}. 

There are two facts crucial for scale-locality of the energy fluxes $\Pi_\ell^{u,b}$. 
First, all the filtered gradient-fields and the central moments can be expressed 
in terms of {\it increments}. In general, for any fields, 
\be \grad\OL{f}_\ell\approx \delta f(\ell)/\ell, \,\,\,\tau_\ell(f,g)
    \approx \delta f(\ell)\delta g(\ell),\,\,\,f_\ell'\approx -\delta f(\ell)
     \lb{inc-def} \ee
where increments are $\delta f(\bx,\br)=f(\bx+\br)-f(\bx),$ $\delta f(\ell)=
\sup_{r<\ell}|\delta f(\br)|,$ and $f_\ell'=f-\OL{f}_\ell$ is the fine-scale (high-pass 
filtered) field. For details, see \cite{Eyink05}.  The second crucial ingredient
for locality is the scaling properties of the increments of velocity and magnetic field:
\be \delta u(\ell) \simeq \ell^{\sigma_u},\,\,\,\,\delta b(\ell)\simeq \ell^{\sigma_b}, \,\,\,\,\,
0<\sigma_{u,b}<1, \lb{scale} \ee
where these relations may be assumed to hold either pointwise, with $\sigma$
the local H\"older exponent, or in the sense of $p$th-order means, $\|\delta f\|_p=
\langle|\delta f(\ell)|^p\rangle^{1/p},$ with $\sigma$ equal to $1/p$ times the 
scaling-exponent $\zeta_p$ of the $p$th-order structure function. As long as 
$0<\sigma_{u,b}<1,$ then (either locally or in the $L_p$-mean sense) the fluxes 
$\Pi_\ell^{u,b}$ are determined by modes all at scales comparable to $\ell$ 
\cite{Eyink05}. For example, the contribution to any increment $\delta f(\ell)$ from scales 
$\Delta\geq\ell$ is represented by $\delta \OL{f}_\Delta(\ell).$ Since the low-pass filtered 
field $\OL{f}_\Delta$ is smooth, its increment may be estimated by Taylor expansion 
and (\ref{inc-def}),(\ref{scale}) as
$$ \delta \OL{f}_\Delta(\ell) \simeq \boell \bdot(\grad\OL{f}_\Delta)
                                                \simeq \ell \Delta^{\sigma-1} 
                                                \simeq \ell^\sigma (\ell/\Delta)^{1-\sigma}, $$ 
and this is negligible for $\Delta\gg\ell$ as long as $\sigma<1.$ On the other hand,
the contribution to any increment $\delta f(\ell)$ from scales $\delta\leq\ell$ is represented 
by $\delta f_\delta'(\ell).$ Since $f'_\delta \approx -(\delta f)(\delta)$ (even without taking 
any difference), (\ref{scale}) implies that
$$ \delta f_\delta'(\ell) \simeq \delta^{\sigma} 
                                        \simeq \ell^\sigma (\delta/\ell)^{\sigma}, $$
and this is negligible for $\delta\ll\ell$ as long as $\sigma>0.$                                          

It is important to emphasize that the scaling laws like (\ref{scale}) used in our proof are 
obtained in all theories of strong MHD turbulence. The Iroshnikov-Kraichnan theory predicts 
that $\sigma_u=\sigma_b=1/4.$ The Goldreich-Sridhar theory predicts distinct scaling for 
increments with displacements in different directions relative to a background field $\bb_0,$
with $\delta u(\ell_\|)\sim \delta b(\ell_\|)\sim \ell_\|^{1/2}$ for displacements in the 
field-parallel direction and $\delta u(\ell_\perp)\sim \delta b(\ell_\perp)\sim \ell_\perp^{1/3}$
for displacements in the perpendicular direction. Such distinctions make no difference 
to our proof, so long as both exponents $\sigma_\|,\,\sigma_\perp$ lie between 0 and 1.
Similarly, our proof is fully compatible with possible intermittency corrections to scaling
exponents. Although the precise scaling of strong MHD turbulence is an open issue,
numerical simulations \cite{Muelleretal03,MininniPouquet09} and natural observations 
\cite{Hily-Blantetal08,Salemetal09} support the validity of the weak condition (\ref{scale}) 
for sufficiently high kinetic and magnetic Reynolds numbers. 

Our arguments imply also the scale-locality of 
cascades of the Els\"asser energies $(1/2)|\bz^\pm|^2,$ with $\bz^\pm=\bu\mp\bb.$
This may be seen by considering the time-derivative of the large-scale 
energy densities $(1/2)|\OL{\bz}_\ell^\pm|^2,$ for which the sink terms are the fluxes 
$-\Pi_\ell^\pm = \grad\OL{\bz}^\pm_\ell\bdots \tau_\ell(\bz^\mp,\bz^\pm) \simeq 
\delta z^\mp(\ell)[\delta z^\pm(\ell)]^2/\ell.$ Since these fluxes are expressed in terms 
of increments, they are scale-local under the weak condition (\ref{scale}). This may 
also be seen from an alternative expression for the Els\"asser energy fluxes which
follow from the Politano-Pouquet relations \cite{PolitanoPouquet98},   $\Pi_\ell^\pm 
=-(3/4\ell)\langle\hat{\boell}\bdot\delta \bz^\mp(\boell)|\delta\bz^\pm(\boell)|^2
\rangle_{{\rm ang}},$ where $\langle\cdot\rangle_{{\rm ang}}$ denotes average over
the displacement directions $\hat{\boell}.$ The scale-locality of cascades of the
Els\"asser energies is particularly important since the foremost phenomenologies
of strong MHD turbulence \cite{Iroshnikov64,Kraichnan65,GoldreichSridhar95, 
Boldyrev05} are based on the picture of counterpropagating Alfv\'en 
wavepackets expressed by the Els\"asser variables $\bz^\pm.$ In terms of these 
variables, the scale-locality properties of MHD turbulence are essentially the same
as those of hydrodynamic turbulence.  
Scale-locality of  the cascades of Els\"asser energies implies scale-locality 
of the flux of cross-helicity $\OL{\bu}_\ell\bdot\OL{\bb}_\ell=(1/4)|\OL{\bz}_\ell^{\,+}|^2
-(1/4)|\OL{\bz}_\ell^{\,-}|^2$ (as well as scale-locality of flux of total energy 
$(1/4)|\OL{\bz}_\ell^{\,+}|^2+(1/4)|\OL{\bz}_\ell^{\,-}|^2$). 

One cascade in MHD turbulence which may be essentially different is that of magnetic 
helicity. The time-derivative of large-scale helicity-density $\OL{\bb}_\ell\bdot\OL{\ba}_\ell$ 
(where $\OL\ba_\ell = (\mbox{curl})^{-1}\OL\bb_\ell$), in addition to space-transport terms, contains  
as a sink term the {\it magnetic helicity flux} $-\Pi_\ell^h=2\OL{\bb}_\ell\bdot\bepsilon_\ell.$ 
Although $\bepsilon_\ell \simeq \delta u(\ell)\delta b(\ell),$ the coarse-grained 
magnetic field $\OL{\bb}_\ell$ will generally be dominated by modes at the forcing 
scale $L.$ 
Thus, magnetic-helicity flux may possibly be dominated 
by non-local triads,  with one mode at the large scale $L.$ 
Similar issues arise for magnetic line-stretching. The time-derivative of large-scale 
kinetic energy $(1/2)|\OL{\bu}_\ell|^2$, in addition to space-transport terms and the 
sink term $-\Pi_\ell^u,$ contains $-\OL{\bb}_\ell^\top\OL{\bS}_\ell\OL{\bb}_\ell$ where
the matrix $\OL{\bS}_\ell=(1/2)[(\grad\OL{\bu}_\ell)+(\grad\OL{\bu}_\ell)^\top]$ is the 
strain from scales $>\ell.$  Likewise, the time-derivative of large-scale magnetic energy 
$(1/2)|\OL{\bb}_\ell|^2$, in addition to space-transport terms and the sink term $-\Pi_\ell^b,$ 
contains $+\OL{\bb}_\ell^\top\OL{\bS}_\ell\OL{\bb}_\ell.$ Thus, this term represents conversion 
between large-scale kinetic and magnetic energy by stretching of coarse-grained field-lines. 
Just as for magnetic helicity flux, this is a ``hybrid'' quantity with  both energy-range and 
inertial-range components. Although $\OL{\bS}_\ell\sim \delta u(\ell)/\ell,$ the coarse-grained
magnetic field $\OL{\bb}_\ell$ can be dominated by modes at scale $L.$ Thus, we cannot 
conclude that  this quantity is dominated by local triads with all modes at scale $\ell.$ 

Indeed, much of the recent discussion about apparent non-locality in MHD turbulence 
has revolved about this conversion term. One of the startling claims that has been 
made in recent numerical studies
\cite{Yousefetal07,Alexakisetal05a,Caratietal06} is that conversion between kinetic and 
magnetic energies proceeds very non-locally, with magnetic modes at scale $\ell$ 
gaining energy equally from all velocity modes at scales $>\ell$ or even predominately
from scale $L\gg \ell.$ In order to examine this claim, we must refine our 
methodology to consider {\it band-pass energies}. Following \cite{EyinkAluie09},
we define pointwise the kinetic and magnetic energy densities in the interval of 
scales  $[\OL\ell,\wt\ell]$ for $\wt\ell > \OL \ell$, as 
$$ e^u_{[\OL\ell,\wt\ell]}=(1/2)\wt{\tau}(\OL{u}_i,\OL{u}_i), \,\,\,\,\,
      e^b_{[\OL\ell,\wt\ell]}= (1/2)\wt{\tau}(\OL{b}_i,\OL{b}_i).  $$
Note that $\OL{(\cdot)}$ now denotes scale $\OL{\ell}$ and $\wt{(\cdot)}$ scale
$\wt{\ell}.$ Their time-derivatives are easily calculated to be
\be \partial_t e^u_{[\OL\ell,\wt\ell]}= 
 -(\wt{\OL{b}_{i}\OL{b}_{j}\OL{S}_{ij}} 
- \wt{\OL{b}}_{i}\wt{\OL{b}}_{j}\wt{\OL{S}}_{ij} )  
+\left(\wt{\OL{\Pi}}^u-\wt{(\OL{\Pi}^u)}\right)+\cdots \lb{Eu} \ee
\be \partial_t e^b_{[\OL\ell,\wt\ell]}= 
 +(\wt{\OL{b}_{i}\OL{b}_{j}\OL{S}_{ij}} 
- \wt{\OL{b}}_{i}\wt{\OL{b}}_{j}\wt{\OL{S}}_{ij} )  
+\left(\wt{\OL{\Pi}}^b-\wt{(\OL{\Pi}^b)}\right)+\cdots \lb{Eb} \ee
where $\cdots$ denotes total divergence terms that correspond to space-transport.
As before, $\OL{\Pi}^u= -\OL{\bS}\bdots\OL{\btau}$ and $\OL{\Pi}^b=-\OL{\bj}\bdot
\OL{\bepsilon},$ and note that the double-filtering length-scale $\wt{\OL{\ell}}
\approx \wt{\ell}$ for $\wt{\ell}\gg \OL{\ell}.$ It is ``obvious'' from these equations 
that the magnetic stretching terms transfer energy between velocity and 
magnetic-field modes only within the same band of length-scales $[\OL\ell,\wt\ell]$.
Clearly, whatever energy is lost or gained from one field by line-stretching reappears 
in or disappears from the other field at the same scale. Non-colliding Alfv\'en waves 
are an example of such non-local triadic exchange which is mediated by a uniform
 magnetic field at the largest scales, but which does not contribute 
 to energy transfer across scales.

Our conclusion above requires some caution, however. A counterexample 
is the Batchelor (viscous-inductive) range that occurs in MHD turbulence with a 
large magnetic Prandtl number $Pr_m=\nu/\eta \gg 1$ 
\cite{Schekochihinetal02}. This range consists of 
length-scales $\ell_\nu \gg \ell \gg \ell_\eta $ far below the inertial-inductive range
$L \gg \ell \gg \ell_\nu,$ with $\ell_\nu$ and $\ell_\eta$ the viscous and resistive
length-scales, resp.  In the Batchelor range, the energy is transferred directly from the 
velocity modes at the viscous scale $\ell_\nu$ into the magnetic-field modes at scales
$\ell\ll \ell_\nu.$ To see that this follows from our eqs. (\ref{Eu})-(\ref{Eb}), we observe 
that the  velocity-gradient in the Batchelor range is almost spatially constant and 
$\grad \OL\bu_\ell \approx \grad\bu$ for all $\ell<\ell_\nu.$ It is thus easy to see that the 
stretching term in (\ref{Eu}) equals $-S_{ij}\wt{\tau}(\OL{b}_i,\OL{b}_j)$ whereas 
the two flux terms become $S_{ij}(\wt{\OL{\tau}(b_i,b_j)} -\wt{\OL{\tau}}(b_i,b_j)).$
(Note that the stress in this range is almost entirely Maxwellian). These terms exactly 
cancel, by the Germano identity \cite{Germano92,MeneveauKatz00}. Thus, 
the line-stretching term acts as an effective source to magnetic energy 
$e^b_{[\OL\ell,\wt\ell]},$ supplied by the flux of kinetic energy directly 
from the viscous scale $\ell_\nu.$

The moral of this 
example is that the energy fluxes also contain line-stretching effects which must be 
considered. Nevertheless, our conclusion is not altered that, {\it in an inertial-inductive
range}, energy conversion by line-stretching is between velocity and magnetic-field modes 
at similar scales. The key point here is the scale-locality of the fluxes, which has already
been established. Because the fluxes only involve modes at comparable scales, they 
cannot transfer energy from very distant scales into scale $\ell$ within an inertial-inductive 
range.  This is not true in the Batchelor range since the velocity field is very 
smooth there ($\sigma_u=1$), violating the condition (\ref{scale}) for scale-locality 
of energy flux in the infrared. 

The studies \cite{Alexakisetal05a,Caratietal06,Yousefetal07}
considered more traditional spectral transfers such as $T_{ub}(K,P)=
\langle \bb^{[P]} (\bb\bdot\grad)\bu^{[K]}\rangle$ and $T_{bu}(K,P)=
\langle \bu^{[P]} (\bb\bdot\grad)\bb^{[K]}\rangle,$ where $\bu^{[K]}$ and
$\bb^{[K]}$ are spectrally band-passed fields for some interval of wavenumbers around
$K.$ Since $T_{ub}(K,P)=-T_{bu}(P,K),$ these can be interpreted (with 
some caution) as energy transfer from the velocity field in band $[K]$ to the 
magnetic field in band $[P].$ Is it possible for the dominant transfers to be between 
distant bands in an inertial-inductive range? The answer is no, if  $[K]$ is a dyadic 
(octave) wavenumber band $[K/2,K].$ It is necessary to use such bands, of equal 
width on a logarithmic scale, in order to permit simultaneous localization of modes
in Fourier and physical space (within the limits of the uncertainty principle). We note
that this is crucial for phenomenological arguments based upon Alfv\'enic wavepackets
with both size and wavenumber specified. The conditions which replace (\ref{scale}) are,
for  $\ba=\bu,\bb$ with $0<\sigma_p^a<1:$
\be \langle|\ba^{[K]}|^p\rangle^{1/p} \simeq K^{-\sigma_p^a}, \,\,\,\,\,
      \langle \big| \grad \ba^{[K]} \big |^{p} \rangle^{1/p} \simeq K^{1-\sigma_p^a}.
\lb{scaling2}\ee
See \cite{AluieEyink09}. If $P<K/2,$ then wavenumber conservation implies that
$T_{ub}(K,P) = -\langle \bu^{[K]}(\bb^{[K/2-P, K+P]}\bdot\grad)\bb^{[P]}\rangle.$
Using the H\"older inequality, this expression is bounded by
$\langle | \grad \bb^{[P]} |^{3} \rangle^{1/3} \langle | \bu^{[K]} |^{3}\rangle^{1/3} 
\langle | \bb^{[K/2-P, K+P]} |^{3}\rangle^{1/3}.$ By (\ref{scaling2})
$$ |T_{ub}(K,P)|\le \const P^{1-\sigma_3^b} K^{-\sigma_3^u -\sigma_3^b} $$
Since $\sigma_3^b < 1$, such transfers for $P \ll K$ are negligible. For
$P>2K,$ $T_{ub}(K,P) = \langle \bb^{[P]}(\bb^{[P/2-K, P+K]}\bdot\grad)\bu^{[K]}\rangle,$
so H\"older inequality and (\ref{scaling2}) imply 
$$ |T_{ub}(K,P)| \le \const K^{1-\sigma_3^u} P^{-2\sigma_3^b}. $$
Since $\sigma_3^b>0,$ transfers for $P \gg K$ are also negligible.

\begin{figure}
\centering
\includegraphics[totalheight=.25\textheight,width=.45\textwidth]{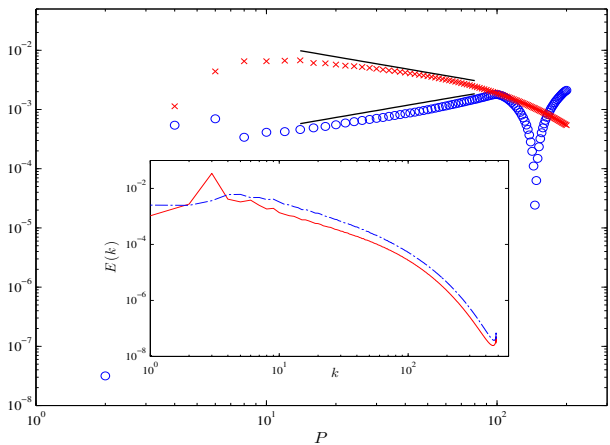}
\caption{ The  transfers $\big|\langle  \partial_j u_i^{[P]} B_i^{[200]} B_j \rangle\big|$ ($\circ$)
and  $\big|\langle  \partial_j u_i^{[4]} B_i^{[P]} B_j \rangle\big|$  ($\times$). Straight lines have 
$\pm2/3$-slopes and extend over the fitting range, which yields a decay rate of $\sim P^{0.68}$ 
for ($\circ$) and $\sim P^{-0.58}$ for ($\times$).
Inset shows velocity (solid line) and magnetic (dashed-dotted line) energy spectra,
which scale close to $E_u\sim E_b\sim k^{-1.61}$ over $k\in[5,80]$.}
 \lb{figIRlocal}
\end{figure}
To test these conclusions numerically we analyze a time snapshot of our 
$1024^3$ MHD simulation in the statistical steady state.  The kinetic and magnetic 
energy spectra of the flow, have a reasonable power-law scaling until around $k=80$
(inset to Fig.\,\ref{figIRlocal}). 
The transfers plotted in Fig.\,\ref{figIRlocal} exhibit off-diagonal ($P\neq K)$ decay 
close to our rigorous upper bounds with exponent $\sigma^b_3\doteq 1/3.$
\begin{figure}
\centering
\includegraphics[totalheight=.15\textheight,width=.3\textwidth]{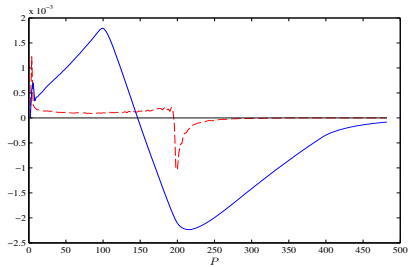}
\caption{ 
The transfers $\langle  \partial_j u_i^{[P/2,P]} B_i^{[100,200]} B_j \rangle$ 
(solid line, same as ($\circ$) plot in Fig.\,\ref{figIRlocal})
and $10^3 \times \langle  \partial_j u_i^{[P-1,P]} B_i^{[199,200]} B_j \rangle$ (dashed line).
The latter is multiplied by $1000$ for comparison.}
 \lb{figLinLog}
\end{figure}
However, the value of this exponent determined from our numerical data (not shown) 
is closer to $\sigma_3^b=1/4,$ consistent with the predictions of \cite{Iroshnikov64,Kraichnan65,Boldyrev05}. 
For this value we obtain rigorous upper bounds $O(P^{0.75})$ for $P\ll K$ and
$O(P^{-.5})$ for $P\gg K,$ which are also close to the observed scaling.   

How are our exact results to be reconciled with the recent numerical studies 
that reach the opposite conclusion? A full discussion is given in our longer
work \cite{AluieEyink_long}, but we make a few remarks here. 
\cite{Schekochihinetal04,Yousefetal07} discussed simulations at lower resolution 
than ours  without carrying out a systematic scaling analysis. 
As for \cite{Caratietal06}, they had an anomalously strong strain at the forcing 
scale $L,$ which can dominate over the local strain at scales $\ell\lesssim L$ 
in an inertial-inductive range of limited extent. 
We also observe this effect over a finite range if we permit such an ``energy 
spike'' at the forcing scale, but it becomes weaker as the amplitude of the spike 
decreases or as the length of the power-law scaling range increases. 
Finally, \cite{Alexakisetal05a} appealed to spectral transfers to justify their 
claim that  the magnetic field at scales $\ell$ in the inertial-inductive range 
receives energy from straining motions at all larger scales $>\ell$, especially from scale 
$L\gg\ell$. However, their DNS study used Fourier bands of linear size $[K-1,K],$ 
which correspond to plane-wave modes which are nonlocalized in space, unlike 
the Alfv\'en wavepackets employed in phenomenological arguments. Such bands
do not properly account for the exponentially growing number of local triads at higher 
wavenumbers, whose aggregate contribution dominates transfers defined with 
logarithmic bands \cite{AluieEyink09}.  Fig.\,\ref{figLinLog}  reproduces the numerical 
result of Fig.\,8 of \cite{Alexakisetal05a} (dashed-line), 
together with our own DNS results using log-bands.  
Clearly, the nonlocal effects observed by \cite{Alexakisetal05a} represent miniscule 
amounts of energy transfer compared with the net contribution of local triads
and  become even smaller as the scale range increases.
In short, the 
numerical results in \cite{Schekochihinetal04,Alexakisetal05a,Caratietal06,Yousefetal07}
do not support any asymptotic non-locality of energy cascade in MHD turbulence.

We thank E. T. Vishniac, S. Chen, M. Wan and D. Shapovalov.  Computer time 
provided by DLMS at the Johns Hopkins University and support from NSF grant 
\# ASE-0428325 are gratefully acknowledged.

\end{document}